\documentclass{article}
\usepackage{fancyhdr}
\usepackage{graphicx} 
\usepackage{hyperref} 
\usepackage{float}
\usepackage{longtable}
\usepackage{colortbl}
\usepackage{xcolor}
\usepackage{adjustbox}
\usepackage{geometry}  
\usepackage{lscape}    
\usepackage{array}
\usepackage{acronym}
\renewcommand{\arraystretch}{1.5} 
\usepackage[backend=biber,style=numeric]{biblatex}
\title{Reporte de Vulnerabilidades en IIoT. Proyecto DEFENDER.}

\author{Pedro Almansa Jiménez\\ Lorenzo Fernández Maimó\\ Ángel Luis Perales Gómez}
\date{Diciembre 2024}
\begin{document}

\maketitle
\thispagestyle{fancy}
\newpage
\tableofcontents
\newpage

\section{Introducci\'on IIoT}

\subsection{Objetivos del Reporte}

El objetivo principal de este informe técnico es realizar un estudio exhaustivo sobre los dispositivos que operan en entornos del Internet Industrial de las Cosas (IIoT), describiendo los escenarios que caracterizan esta categoría y analizando las vulnerabilidades que comprometen su seguridad. Para ello, se busca identificar y analizar las principales clases de dispositivos IIoT, describiendo sus características, funcionalidades y roles dentro de los sistemas industriales. Este análisis permitirá comprender cómo estos dispositivos interactúan y cumplen con los requerimientos de entornos industriales críticos.\\

Asimismo, el informe examina los entornos específicos en los que operan estos dispositivos, destacando las particularidades de los escenarios industriales y las condiciones bajo las cuales funcionan. Además, se analizan las vulnerabilidades exponiendo los vectores, objetivos, impacto y consecuencia de las mismas. Posteriormente se explican las fases típicas de un ataques junto a una selección de casos reales de ataques documentados junto con su clasificación según la taxonomía expuesta en el apartado 3. Esto proporciona una visión integral de las posibles amenazas que comprometen la seguridad, evaluando el impacto que estas vulnerabilidades pueden tener en los entornos industriales.\\

Finalmente, se presentan una recopilación de algunas de las contra medidas de seguridad mas recientes y eficaces como soluciones a los problemas de seguridad de los sistemas industriales. Hace especial énfasis en enfocar la importancia del Machine Learning en el desarrollo de estos enfoques. 

\subsection{Definición y Características}

El Internet Industrial de las Cosas (IIoT, por sus siglas en inglés) es una extensión especializada del Internet de las Cosas (IoT), enfocada en aplicaciones industriales y sistemas ciberfísicos. Mientras que el IoT conecta dispositivos de uso cotidiano como electrodomésticos, sensores domésticos y wearables, el IIoT se centra en la interconexión de máquinas, sensores avanzados, actuadores y sistemas industriales en entornos como fábricas, plantas energéticas y cadenas de suministro. Esta tecnología busca no solo recopilar y analizar datos, sino también optimizar procesos industriales críticos mediante la automatización.\\

Entre las principales características del IIoT destaca su alta conectividad y escalabilidad, diseñada para manejar una enorme cantidad de dispositivos y datos en tiempo real. A diferencia del IoT, el IIoT requiere protocolos de comunicación robustos y confiables que garanticen la integridad y disponibilidad de los datos en entornos industriales exigentes.\\

Otra característica clave del IIoT es su enfoque en la resiliencia y seguridad, dado que los dispositivos suelen operar en infraestructuras críticas donde cualquier vulnerabilidad puede tener consecuencias graves, como interrupciones operativas o riesgos para la seguridad humana hacen que el trabajo y la mejora de métodos y herramientas para mitigar estas amenazas se haya vuelto algo crucial.\\

Por último, el IIoT se caracteriza por su interoperabilidad y capacidad de integración con tecnologías emergentes como la computación en la nube, el edge computing y los sistemas de inteligencia artificial. Esto permite que los dispositivos no solo interactúen entre sí, sino que también integren los datos recopilados en plataformas de análisis para mejorar la toma de decisiones, reducir costos operativos y aumentar la eficiencia de las operaciones industriales.

\subsection{Aplicaciones Clave}

El Internet Industrial de las Cosas (IIoT) ha revolucionado múltiples industrias al introducir dispositivos conectados que mejoran la eficiencia y automatizan procesos volviéndose un elemento prácticamente imprescindible en multitud de sectores críticos. A continuación, se destacan las aplicaciones clave y los sectores donde el IIoT está teniendo un rol cada vez mayor:

\begin{itemize}
    \item \textbf{Industrial y manufacturero}: El IIoT desempeña un papel fundamental en la automatización de líneas de producción, el mantenimiento predictivo de maquinaria y la optimización de la logística. Los sensores y dispositivos conectados recopilan datos en tiempo real, permitiendo la detección temprana de fallos, la reducción de tiempos de inactividad y el incremento de la productividad mediante análisis avanzados y sistemas de control.

    \item \textbf{Energía e infraestructuras críticas}: El IIoT se aplica para la monitorización de redes eléctricas, plantas de energía y sistemas de distribución. Dispositivos conectados permiten gestionar de manera eficiente el consumo energético, predecir fallos en infraestructuras como turbinas eólicas o redes de distribución, y garantizar la seguridad operativa en plantas nucleares o de petróleo y gas.

    \item \textbf{Transporte y logística}: El IIoT facilita la gestión de flotas mediante el rastreo de vehículos en tiempo real, la monitorización de condiciones de transporte de mercancías sensibles (como alimentos perecederos o productos farmacéuticos) y la optimización de rutas logísticas. Esto se traduce en menores costos operativos y mayor confiabilidad en las cadenas de suministro.

    \item \textbf{Salud y medicina industrial}: El IIoT se utiliza en hospitales inteligentes y plantas farmacéuticas, donde sensores y dispositivos conectados garantizan el monitoreo de equipos médicos, el control de condiciones ambientales en laboratorios y la seguridad en procesos de producción farmacéutica, cumpliendo con estrictos estándares regulatorios.
\end{itemize}

\subsection{Crecimiento del Mercado y su Impacto en la Seguridad}
El mercado de IoT, y en particular el de IIoT, ha experimentado un crecimiento exponencial en los \'ultimos a\~nos, impulsado por la creciente adopci\'on de tecnolog\'ias digitales y la necesidad de soluciones inteligentes en entornos industriales cr\'iticos. El tama\~no del mercado de Internet de las cosas se estima en USD 1,17 billones en 2024 y se espera que alcance los USD 2,37 billones para 2029.\footnote{Fuente: Statista Research Department, 2024. Disponible en: \url{https://es.statista.com/temas/6976/el-internet-de-las-cosas-iot/}}. Seg\'un informes recientes, se espera que el n\'umero de dispositivos IoT activos supere los 30 mil millones para 2025, con una proporci\'on significativa atribuida a dispositivos IIoT.\\

\begin{figure}[h!]
    \centering
    \includegraphics[width=0.7\textwidth]{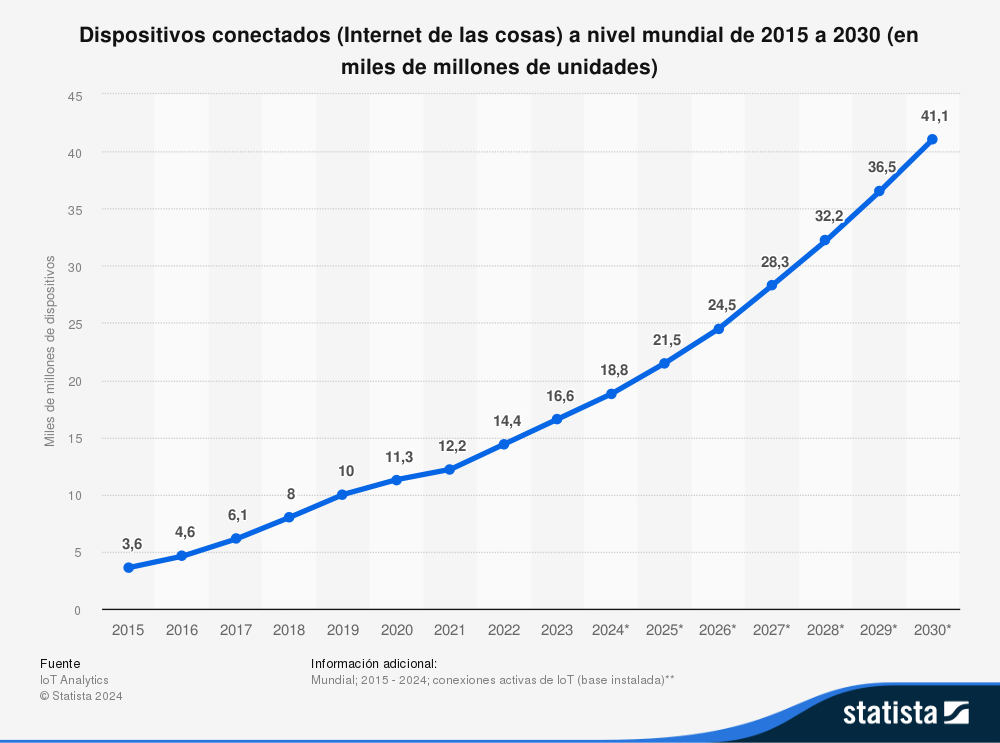}
    \caption{Proyecci\'on del crecimiento del mercado de IoT hasta 2030.}
    \label{fig:crecimiento_iot}
\end{figure}

Dentro de la industria se espera que los avances en dispositivos de campo, sensores y robots ampl\'ien a\'un m\'as el alcance del mercado, especialmente en aplicaciones de la Industria 4.0. Las tecnolog\'ias IIoT est\'an superando la escasez de mano de obra en el sector manufacturero y mejorando la eficiencia en la producci\'on. Para cada vez m\'as organizaciones, el uso de tecnolog\'ias como la robotizaci\'on y el mantenimiento predictivo forma parte de las operaciones diarias. Un estudio realizado por la empresa de IoT Industrial (IIoT) Microsoft Corporation encontr\'o que el 85\% de las empresas tienen al menos un proyecto de caso de uso de IIoT.\footnote{Fuente: Microsoft Corporation. Disponible en: \url{https://news.microsoft.com/es-es/2019/08/06/microsoft-presenta-iot-signals-un-estudio-sobre-el-estado-de-adopcion-del-internet-of-things/}.}\\

Sin embargo, este crecimiento tambi\'en ha expuesto importantes retos de seguridad, particularmente en el \'ambito de IIoT, donde los dispositivos forman parte de infraestructuras cr\'iticas. Muchos dispositivos carecen de dise\~nos seguros, lo que los hace vulnerables a ataques como el acceso no autorizado, la manipulaci\'on de datos y el uso indebido para redes de botnets. La seguridad de los dispositivos IIoT no solo afecta a los usuarios finales, sino que tambi\'en tiene implicaciones significativas para la infraestructura cr\'itica y la econom\'ia global. Un informe de Kaspersky destaca que los ataques de malware dirigidos a dispositivos IoT aumentaron significativamente en la primera mitad de 2023 en comparaci\'on con 2022.\footnote{Fuente: Kaspersky Lab, "IoT under attack: 400\% increase in malware targeting connected devices", 2023. Disponible en: \url{https://www.kaspersky.com/}.}\\

\newpage

\section{Entorno IIoT}

Los entornos del \textbf{Internet Industrial de las Cosas (IIoT)} son sistemas complejos que combinan dispositivos físicos, software y redes avanzadas para operar en escenarios industriales críticos. A diferencia de las aplicaciones genéricas del IoT, los entornos IIoT requieren una infraestructura robusta y específica, diseñada para cumplir con las exigencias de sectores como la manufactura, la energía y el transporte. \\

\begin{figure}[H]
    \centering
    \includegraphics[width=1.0\textwidth]{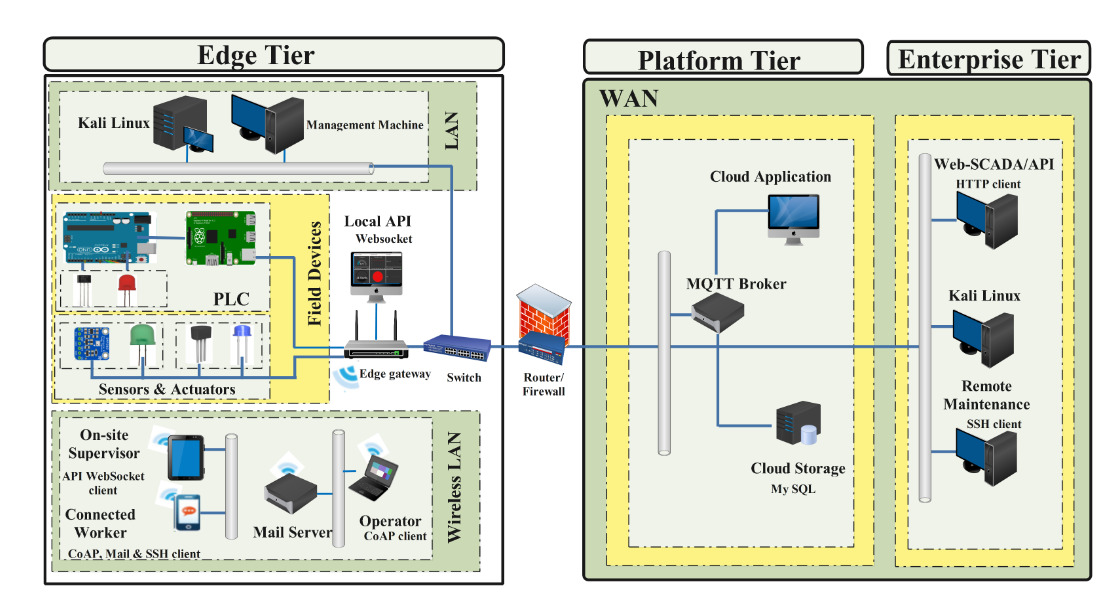}
    \caption{Entorno IIoT. Imagen adaptada de~\cite{xiiotid}.}
    \label{fig:entorno_iiot}
\end{figure}

Un entorno IIoT típico, como se ilustra en la Figura~\ref{fig:entorno_iiot}, integra múltiples niveles, desde los dispositivos de campo, como sensores y actuadores, hasta plataformas avanzadas para análisis de datos y gestión en la nube. Esta estructura modular permite la interoperabilidad entre diferentes componentes y sistemas, algo crucial en un entorno tan diverso.\\

Además, estos entornos dependen de protocolos de comunicación industriales para asegurar la transmisión eficiente y segura de datos entre dispositivos y plataformas. La integración de métricas relevantes, como la latencia de la red, la disponibilidad del sistema y la seguridad de los datos, es esencial para medir el rendimiento y garantizar la sostenibilidad de las operaciones. Estas características hacen que los entornos IIoT sean fundamentales para la transformación digital en la industria, sentando las bases para la Industria 4.0.\\

En este sentido, la implementación de tecnologías de comunicación de vanguardia, como el 5G, cobra especial relevancia, ya que no solo reduce la latencia a niveles mínimos al facilitar respuestas casi en tiempo real, sino que también ofrece mayor ancho de banda y confiabilidad. Esto resulta fundamental para afianzar la coordinación y eficiencia de los procesos industriales en un entorno cada vez más interconectado.\\

\newpage
\subsection{Dispositivos IIoT}

Los dispositivos del Internet Industrial de las Cosas (IIoT) se pueden clasificar en diversas categorías, cada una con funciones específicas que contribuyen a la automatización, el monitoreo y la optimización de procesos en entornos industriales. A continuación, se presentan algunas de las principales categorías y sus dispositivos más destacados:

\subsubsection{Dispositivos de Adquisición de Datos}

\textbf{Sensores Industriales:} Son dispositivos esenciales para recopilar datos en tiempo real de variables como temperatura, presión, vibración, humedad, entre otros. Su precisión y capacidad de resistencia a condiciones extremas los hacen ideales para aplicaciones críticas en plantas industriales.\\

\textbf{Cámaras y Sensores de Visión:} Utilizados para inspección visual y control de calidad, estos dispositivos incorporan capacidades de inteligencia artificial para identificar defectos, realizar mediciones y optimizar procesos. Las cámaras de visión infrarroja, por ejemplo, son fundamentales para detectar anomalías térmicas.

\begin{figure}[H]
    \centering
    \includegraphics[width=0.45\textwidth]{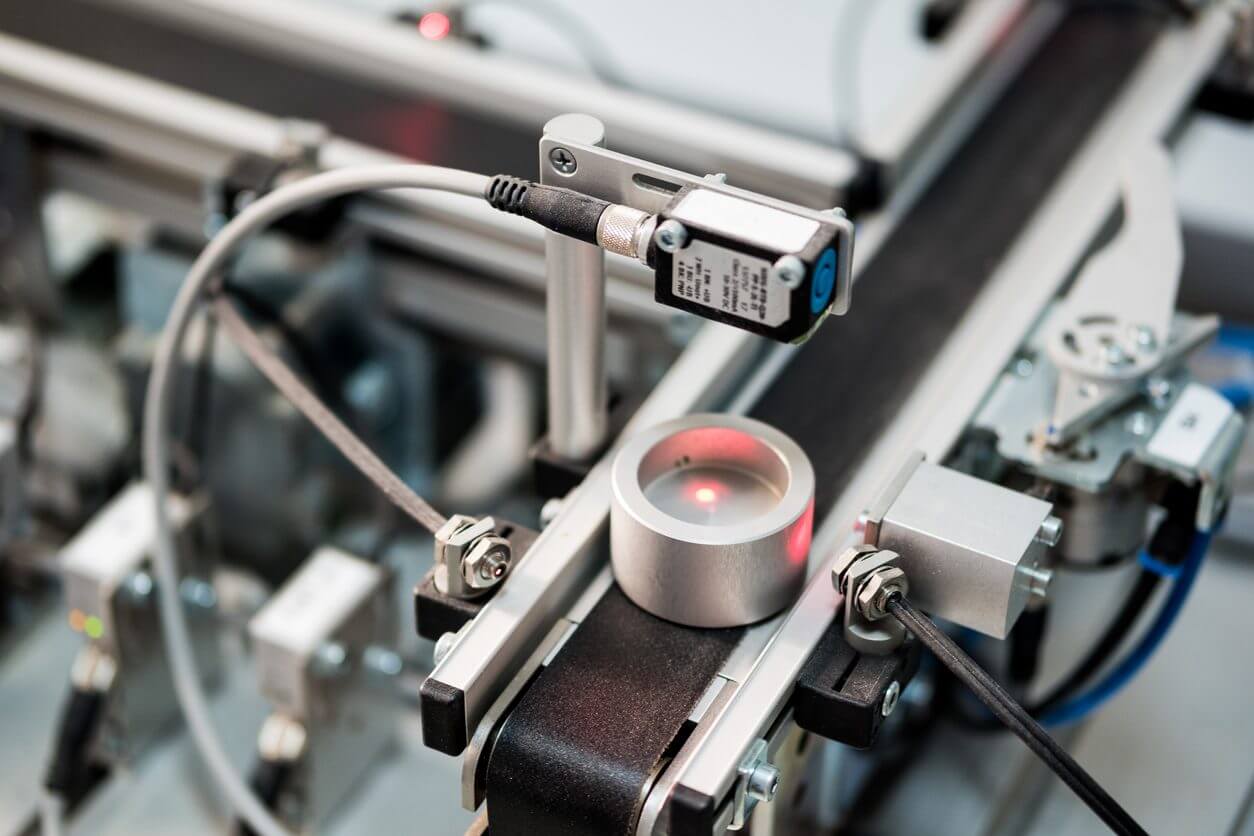}
    \includegraphics[width=0.35\textwidth]{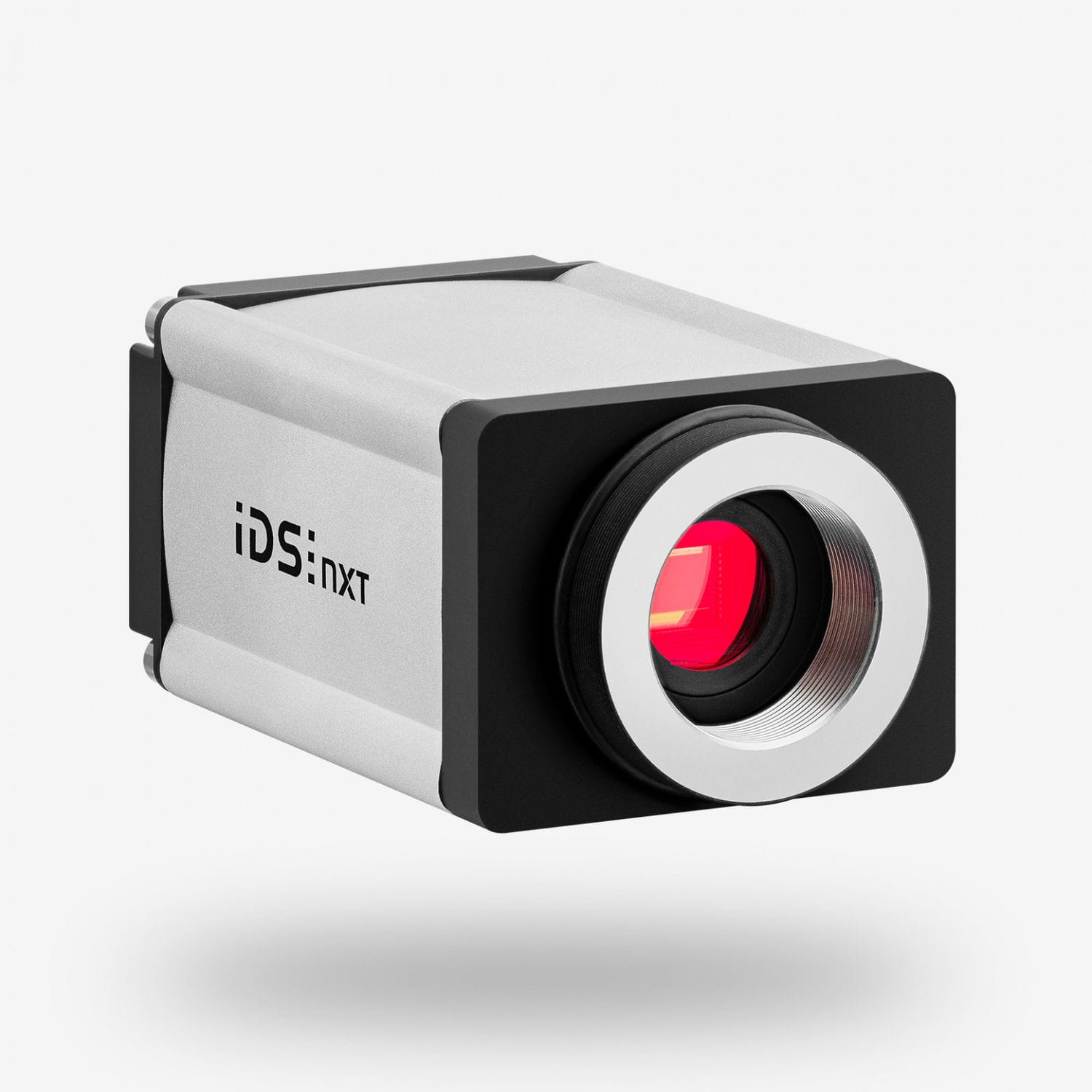}
    \caption{Izquierda: Sensor industrial. Derecha: Cámara de visión infrarroja.}
    \label{fig:adquisicion_datos}
\end{figure}

\subsubsection{Dispositivos de Control}

\textbf{Controladores Lógicos Programables (PLCs):} Estos equipos gestionan procesos automatizados, tomando decisiones en función de datos recopilados por sensores. Su programación flexible permite adaptarlos a diversos procesos industriales, desde líneas de ensamblaje hasta sistemas de distribución.\\

\begin{figure}[H]
    \centering
    \includegraphics[width=0.55\textwidth]{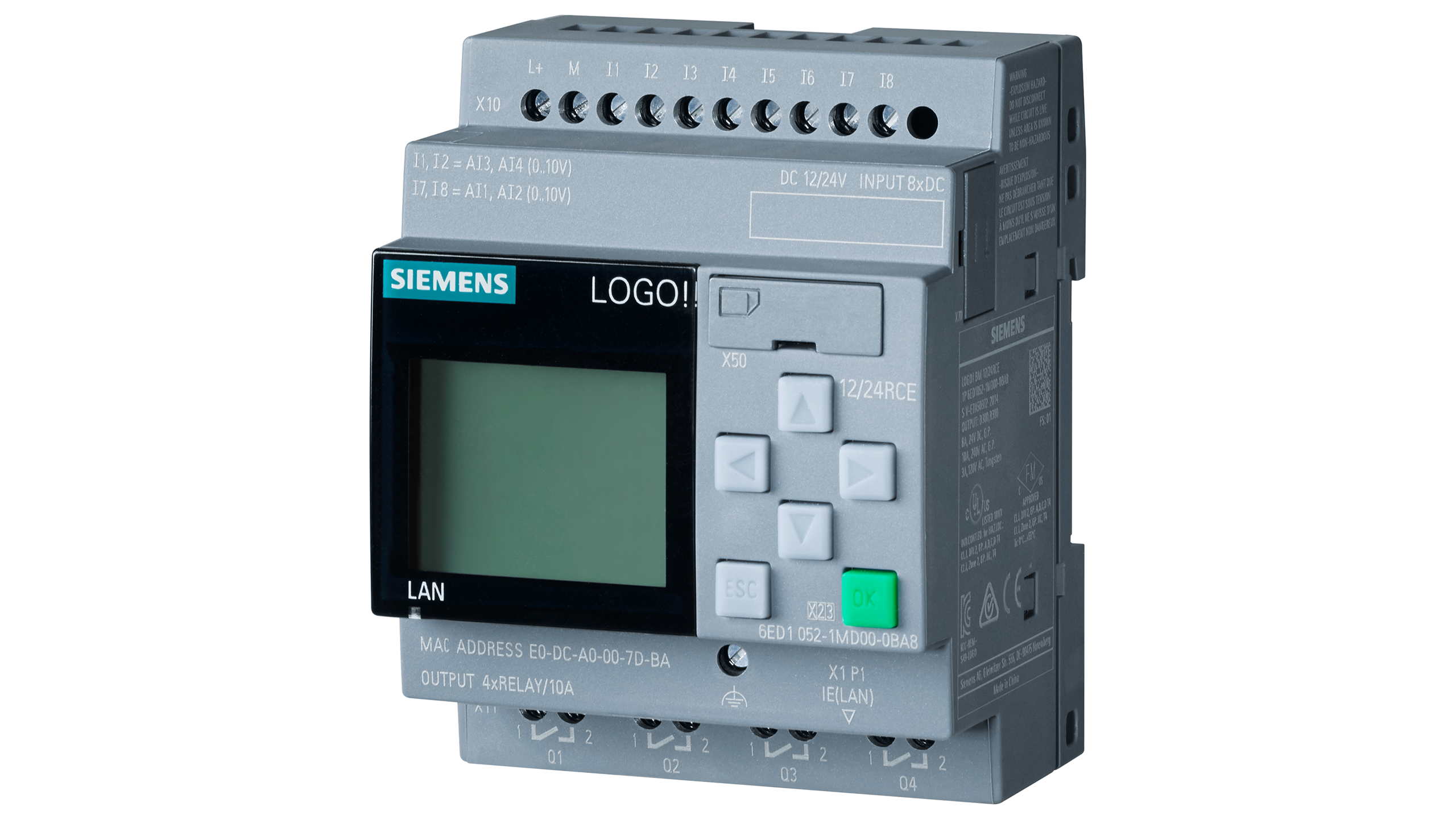}
    \caption{Izquierda: Controlador Lógico Programable (PLC).}
    \label{fig:control_automatizacion}
\end{figure}
\newpage
\subsubsection{Dispositivos de Conexión y Gestión de Redes}

\textbf{Gateways Industriales:} Actúan como intermediarios entre redes locales y servicios en la nube. Facilitan la transmisión de datos a plataformas de análisis, integrando dispositivos de campo con sistemas empresariales para monitoreo remoto y gestión centralizada.\\

\textbf{Mail Servers y Redes LAN:} En entornos IIoT, servidores locales y gateways aseguran la comunicación eficiente entre niveles del sistema. Esto incluye tanto conexiones a dispositivos como la transmisión segura de datos hacia servidores remotos o aplicaciones en la nube.

\begin{figure}[H]
    \centering
    \includegraphics[width=0.45\textwidth]{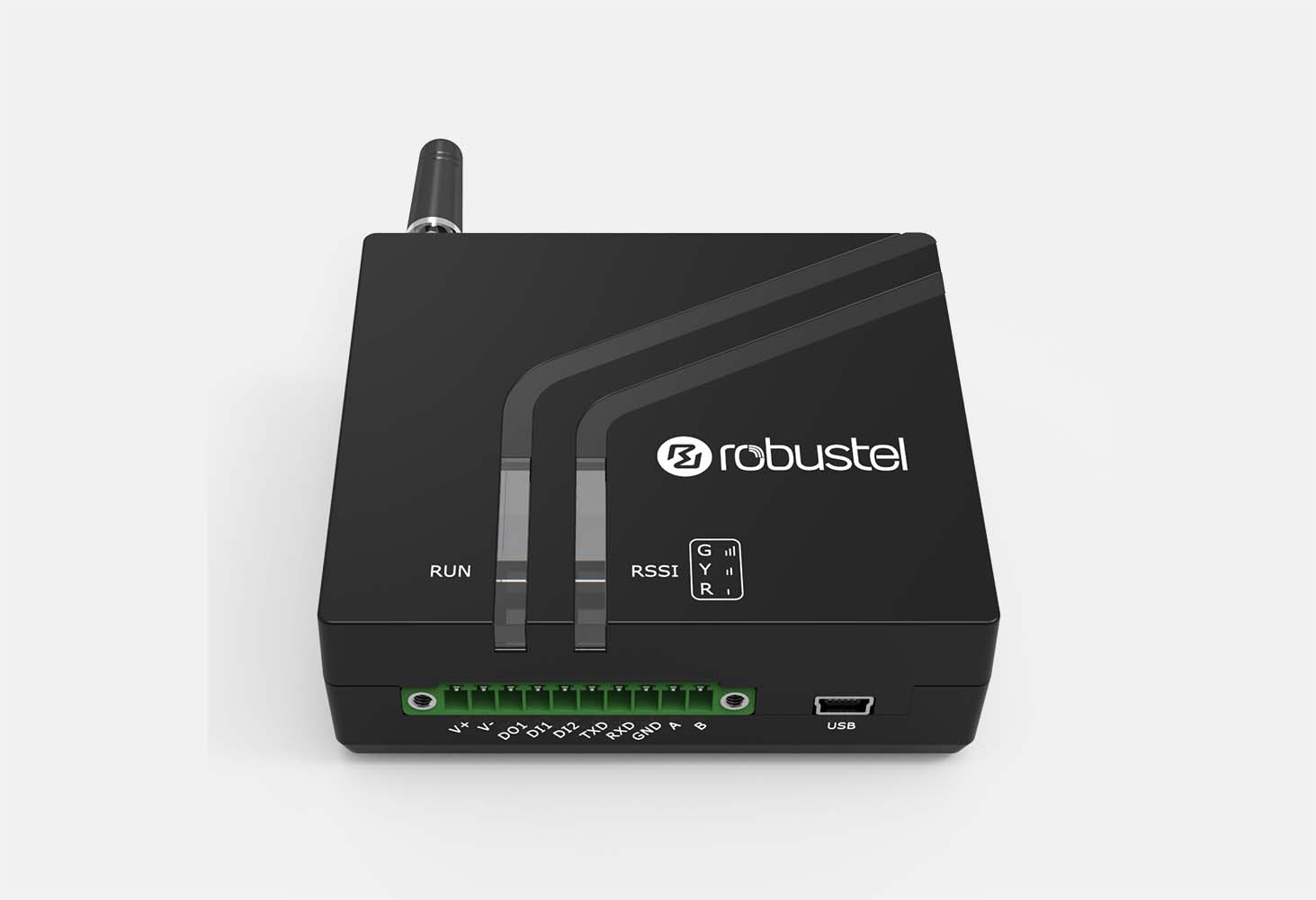}
    \caption{Gateway Industrial. }
    \label{fig:conexion_gestion_redes}
\end{figure}

\subsubsection{Sistemas de Automatización}

\textbf{Actuadores:} Convierte señales electrónicas en acciones físicas, como abrir válvulas, mover piezas o accionar motores. Los actuadores pueden ser hidráulicos, neumáticos o eléctricos, dependiendo de las necesidades del entorno.\\

\textbf{Robots Industriales:} Diseñados para realizar tareas específicas como ensamblaje, soldadura o transporte de materiales. Los robots multiarticulados son altamente versátiles y mejoran significativamente la precisión y la eficiencia en entornos industriales.

\begin{figure}[H]
    \centering
    \includegraphics[width=0.45\textwidth]{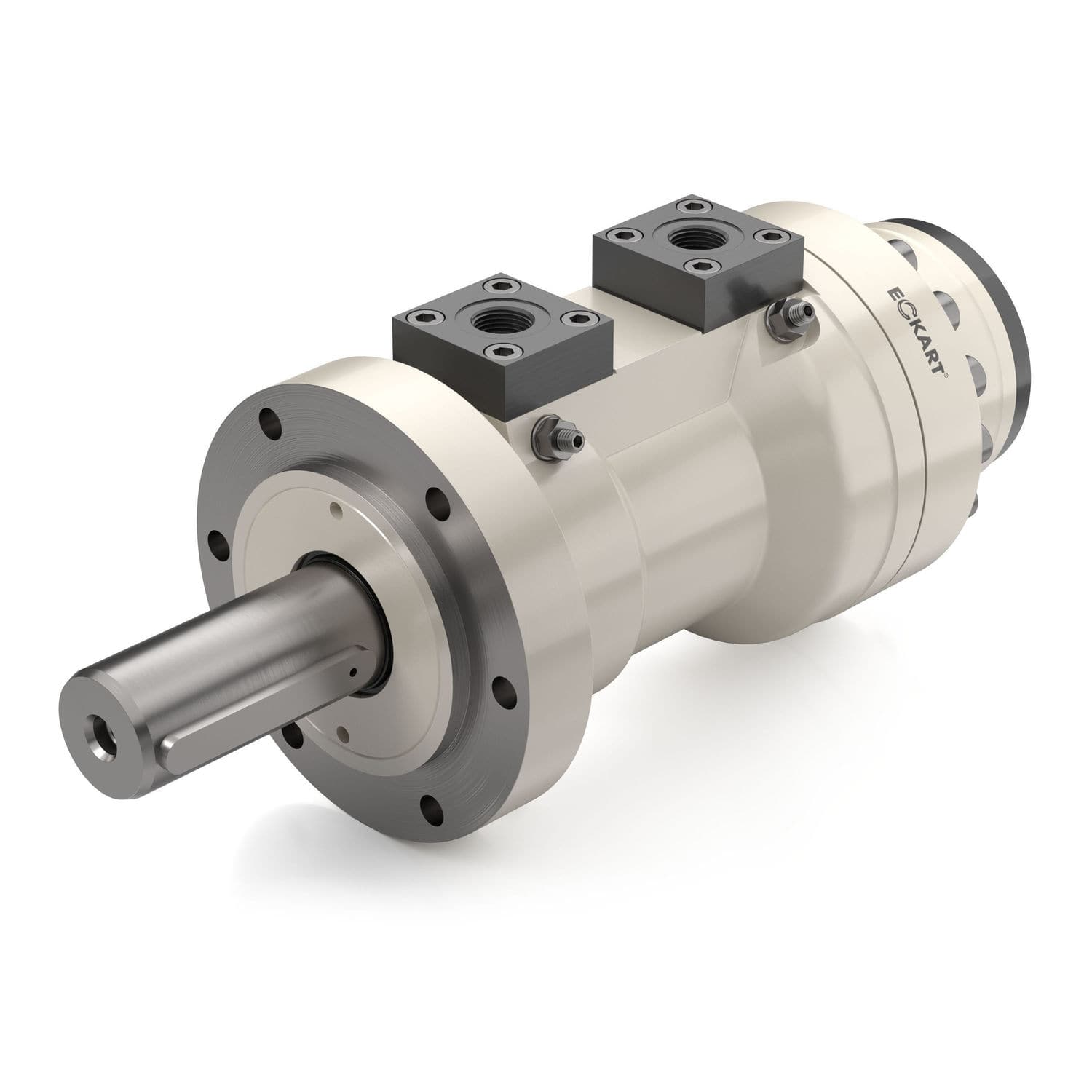}
    \includegraphics[width=0.45\textwidth]{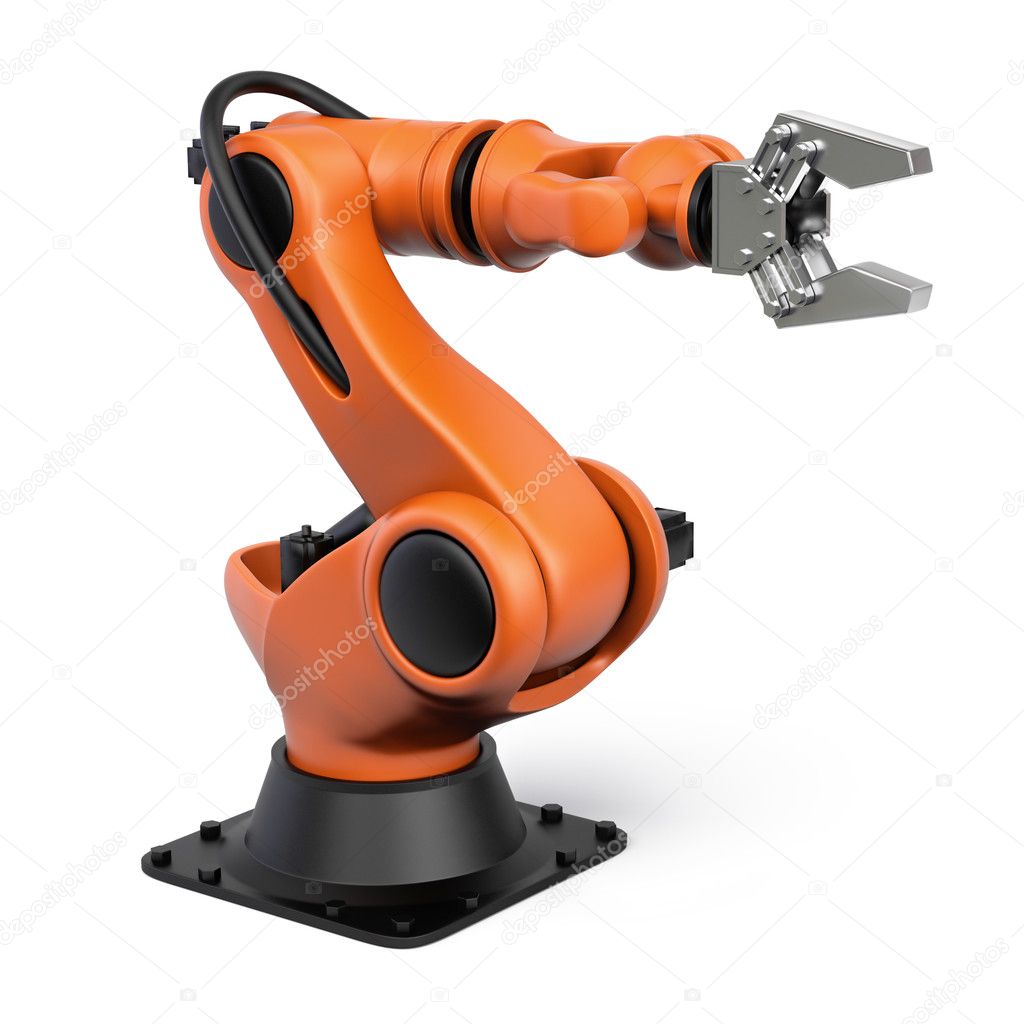}
    \caption{ Izquierda: Actuador hidráulico industrial. Derecha: Brazo robótico multiarticulado. }
    \label{fig:sistemas_avanzados}
\end{figure}
\newpage

\subsection{Arquitectura del Entorno}

La arquitectura de un entorno IIoT puede ser tremendamente diversa. Cada sector industrial, cada conjunto de dispositivos y cada estrategia de negocio demandan requisitos específicos que pueden variar considerablemente. Elementos como el tamaño de la empresa o fábrica, o el presupuesto disponible para invertir en infraestructura IIoT, marcan factores diferenciales a la hora de moldear los entornos. Esto, a su vez, da lugar a la existencia de entornos "híbridos", en los que elementos de IIoT se combinan con máquinas y sistemas desfasados y obsoletos, lo que no hace sino añadir una capa más de complejidad a la defensa de estos entornos.\\

Sin embargo para para tener un modelo de referencia hemos seleccionado el fundamentado en el modelo de referencia \textit{Industrial Internet Reference Architecture (IIRA)}~\cite{iira}, establecido por el \textit{Industrial Internet Consortium (IIC)}~\cite{iic}.
. Este modelo proporciona un marco estandarizado para diseñar sistemas IIoT que integren de manera eficiente los flujos de datos, las interacciones entre dispositivos y las capas de control necesarias en entornos industriales críticos. La estructura propuesta comprende tres niveles principales, cada uno con funciones específicas para garantizar interoperabilidad, escalabilidad y seguridad:

\begin{itemize}
    \item \textbf{Capa de Borde (\textit{Edge Tier}):}
    Este nivel constituye la interfaz directa entre los procesos físicos y el sistema digital. Los \textbf{sensores y actuadores} son responsables de monitorear y controlar variables críticas, como temperatura, presión y vibración, en tiempo real. Los \textbf{controladores lógicos programables (PLCs)}, ampliamente utilizados en sistemas industriales, gestionan los procesos automatizados mediante lógica predefinida. Por último, los \textbf{gateways industriales} actúan como puntos de convergencia, conectando la red local de dispositivos (OT) con servicios en la nube o plataformas (IT), utilizando protocolos estándar como OPC UA y MQTT. Esta capa se diseña para garantizar confiabilidad, baja latencia y adaptabilidad a entornos dinámicos.

    \item \textbf{Capa de Plataforma (\textit{Platform Tier}):}
    Este nivel opera como el núcleo del procesamiento y almacenamiento de datos, gestionando tanto información en tiempo real como datos históricos. Las \textbf{bases de datos y sistemas de almacenamiento} soportan grandes volúmenes de datos generados por los dispositivos de la capa de borde. Además, esta capa incluye \textbf{sistemas de análisis avanzados}, que permiten extraer patrones, detectar anomalías y ejecutar algoritmos de inteligencia artificial o aprendizaje automático. El diseño de esta capa sigue las recomendaciones del IIRA para garantizar interoperabilidad y soporte a aplicaciones de alto nivel.

    \item \textbf{Capa Empresarial (\textit{Enterprise Tier}):}
    Diseñada para la interacción humana y la toma de decisiones estratégicas, esta capa incluye herramientas como \textbf{sistemas SCADA y aplicaciones API} para monitorear y controlar procesos en tiempo real. Además, integra \textbf{sistemas de gestión empresarial (ERP)} que facilitan la planificación, supervisión y optimización de operaciones. Según el IIRA, esta capa debe priorizar la seguridad y la accesibilidad, permitiendo que los datos y análisis de las capas inferiores respalden decisiones empresariales informadas.

\end{itemize}

Si bien la adopción de este estándar como estructura base de una arquitectura IIoT facilita la labor de estudio de entornos de esta clase, cabe recordar que los entornos IIoT puede ser tremendamente diversos. Sin embargo es útil para situarnos en un marco ampliamente aceptado de lo que debería ser un escenario de este estilo.\\\\

\subsection{Protocolos y Comunicaciones}
\label{subsec:protocolos_comunicaciones}

La arquitectura típica de un entorno \Acp{IIoT} integra distintos niveles o capas de comunicación: dispositivos (sensores, actuadores, cámaras), pasarelas (\textit{gateways}), sistemas de gestión de datos y plataformas en la nube o \textit{edge computing} \cite{mekala_cybersecurity_iiot}. Cada una de estas capas se apoya en diferentes protocolos de red, concebidos para satisfacer requisitos de baja latencia, escalabilidad, capacidad de transmisión en tiempo real y compatibilidad con dispositivos de recursos reducidos. Según el análisis realizado por Ramírez Delgado y Díaz-Piraquive \cite{ramirez_protocolo_iot} y el estudio de Mekala et al.\ \cite{mekala_cybersecurity_iiot}, la adopción de protocolos de comunicación para entornos IIoT se concentra, principalmente, en los siguientes conjuntos:

\begin{itemize}
    \item \textbf{Protocolos M2M (Machine-to-Machine).}
    Estos protocolos gestionan el intercambio directo de datos entre equipos industriales (p.\,ej., sensores, \acp{PLC}, actuadores). Algunos ejemplos destacados incluyen:

    \begin{itemize}
        \item \textbf{MQTT (Message Queue Telemetry Transport).} 
        Protocolo ligero de mensajería basado en \acs{TCP}/\acs{IP}, con un modelo de suscripción y publicación (\textit{publish--subscribe}). Empleado en múltiples sectores industriales gracias a su facilidad de configuración y bajo consumo de ancho de banda.    
        \item \textbf{CoAP (Constrained Application Protocol).} 
        Diseñado para dispositivos con recursos limitados y enfocado en entornos \ac{IoT} basados en \acs{IPv6}. Ofrece modos de comunicación \textit{request--response} y \textit{publish--subscribe}, con un encabezado reducido que facilita la interacción en redes de baja potencia.
        \item \textbf{DDS (Data Distribution Service).} 
        Enfatiza un modelo de datos centrado en la comunicación entre nodos de \textit{edge}. Es descentralizado y soporta el envío de datos tanto por \acs{UDP}/\acs{IP} como por \acs{TCP}/\acs{IP}. Se utiliza a menudo en sistemas de control distribuidos que requieren un bajo retardo y gran robustez (p.\,ej., gestión de redes eléctricas, tráfico aéreo o transporte).
        \item \textbf{AMQP (Advanced Message Queuing Protocol).}
        Protocolizado en un modelo de \textit{publish--subscribe} confiable y orientado a mensajería corporativa. Facilita el intercambio fiable de datos con acuses de recibo y es común en sistemas donde se requiere transaccionalidad y asincronía.
        \item \textbf{MODBUS/TCP.}
        Utilizado profusamente en redes industriales para la comunicación entre dispositivos de supervisión (\ac{SCADA}) y \acp{PLC}. Emplea un esquema maestro--esclavo sobre \acs{TCP}/\acs{IP} y destaca por su facilidad de implementación y por su amplia adopción en la industria.
        \item \textbf{CAN (Controller Area Network).} 
        Orientado a entornos con requisitos de robustez y operación en tiempo real (por ejemplo, vehículos o maquinaria industrial). Opera sobre un bus de comunicaciones en serie, optimizado para escenarios de alta fiabilidad.
        \item \textbf{WirelessHART.}
        Estándar abierto ideado para la comunicación inalámbrica en procesos industriales. Emplea topologías en malla (\textit{mesh}) y habilita la conectividad entre sensores y actuadores en condiciones difíciles, aportando alta disponibilidad.
        \item \textbf{NB-IoT (Narrowband IoT).} 
        Tecnología \ac{LPWAN} celular de bajo consumo y gran cobertura, pensada para la conexión de una gran cantidad de dispositivos con un coste de mantenimiento reducido (por ejemplo, \textit{smart buildings}, \textit{smart cities}, etc.).
    \end{itemize}

    \item \textbf{Protocolos H2M (Human-to-Machine).}
    Estos protocolos permiten la interacción con los operadores humanos para fines de monitorización y control:
    \begin{itemize}
        \item \textbf{HTTP (HyperText Transfer Protocol).} 
        Presente en interfaces web para la administración de dispositivos y la visualización de datos. En el ámbito IIoT, se utiliza a menudo en plataformas de supervisión remota y herramientas de diagnóstico.
        \item \textbf{CoAP.} 
        Además de su uso en M2M, sus modos de operación ligeros también pueden facilitar la comunicación con aplicaciones humanas, sobre todo en redes con recursos muy restringidos.
    \end{itemize}

    \item \textbf{Protocolos de Red y Estructuras de Enlace.}
    \begin{itemize}
        \item \textbf{\acs{TCP}/\acs{IP} y \acs{UDP}.} 
        Forman la base del ecosistema IIoT al proporcionar transporte confiable o de baja latencia, respectivamente.
        \item \textbf{Fieldbus (p.\,ej., PROFIBUS, DeviceNet, Foundation Fieldbus).}
        Agrupa diversos estándares para la automatización industrial, interconectando sensores, actuadores y \acp{HMI} en topologías como buses, anillos o mallas. Permite la digitalización de procesos críticos y la integración con capas de supervisión.
        \item \textbf{LoRaWAN (Long Range Wide Area Network).} 
        Aunque no se detalla profundamente en algunos estudios de IIoT industriales, es considerado un protocolo \ac{LPWAN} relevante para la monitorización de sensores distribuidos a gran distancia y con muy bajo consumo.
        \item \textbf{ZigBee y 6LoWPAN.} 
        Orientados a redes de baja potencia y bajo costo. Suelen utilizarse en escenarios de corto alcance (ZigBee) o para habilitar conectividad \acs{IPv6} en dispositivos con recursos muy limitados (6LoWPAN).
    \end{itemize}

\end{itemize}

Como señalan Qiu et al.\ \cite{qiu_edge_computing} y Boyes et al.\ \cite{boyes_iiot_framework}, la adopción de estos protocolos permite a la industria abordar el reto de la interconexión masiva de equipos heterogéneos, habilitando la monitorización en tiempo real y la integración de datos para su análisis en la nube. Cada protocolo presenta capacidades específicas (estructura de mensajes, modelo de comunicación, escalabilidad), y su elección depende de factores como el número de dispositivos, los requerimientos de latencia o la disponibilidad de ancho de banda. La compatibilidad entre capas de red y la correcta configuración de los dispositivos resultan, por tanto, fundamentales para asegurar la fluidez y fiabilidad de las comunicaciones en un entorno industrial de nueva generación.
\newpage

\subsection{Controladores en el Entorno Industrial}

Los sistemas de control son componentes esenciales en los entornos industriales, desempeñando un papel crítico en la operación, supervisión y optimización de infraestructuras clave como fábricas inteligentes, cadenas de suministro y procesos mineros. Estos sistemas permiten monitorear variables físicas, gestionar activos distribuidos y garantizar la eficiencia operativa mediante la interacción entre sensores, actuadores y controladores. Su función principal radica en recopilar datos en tiempo real, analizarlos y ejecutar acciones correctivas o preventivas para mantener los procesos dentro de los parámetros deseados. Los sistemas de control pueden adoptar diferentes arquitecturas dependiendo de las necesidades del entorno, destacando los sistemas centralizados, descentralizados y jerárquicos como los modelos más comunes~\cite{xu2018survey}. A continuación, se describen las características principales de cada uno de estos enfoques.

\begin{itemize}
    \item \textbf{Sistemas de Control Centralizado:} En los sistemas de control centralizado, un único controlador supervisa y gestiona múltiples subsistemas dentro de la arquitectura del sistema. Este controlador centralizado recibe datos de sensores que registran el estado operativo de los subsistemas y envía señales de comando a los actuadores correspondientes para ajustar su comportamiento. Un ejemplo clásico de este tipo de sistema es el \textit{SCADA} (Sistema de Control Supervisado y Adquisición de Datos), ampliamente utilizado en redes eléctricas y sistemas de distribución de agua~\cite{khujamatov2021iot}. Aunque estos sistemas ofrecen ventajas como la centralización de datos y la facilidad de supervisión, también presentan vulnerabilidades significativas debido a su dependencia de plataformas heredadas y la falta de actualizaciones frecuentes, lo que los expone a amenazas cibernéticas~\cite{yadav2021scada}.

    \item \textbf{Sistemas de Control Descentralizado:} Por otro lado, los sistemas de control descentralizado distribuyen la responsabilidad del control entre varios controladores individuales, cada uno dedicado a un subsistema específico. Esta arquitectura permite una mayor flexibilidad y capacidad de respuesta local, ya que los controladores no dependen de un único punto de control. Un ejemplo destacado es el \textit{Sistema de Control Distribuido} (DCS), que combina múltiples controladores para coordinar procesos complejos de producción industrial~\cite{xu2018survey}. Otro caso relevante son los \textit{Controladores Lógicos Programables} (PLCs), que interpretan señales de sensores y generan respuestas automáticas a intervalos regulares. Estos sistemas son ampliamente utilizados en industrias manufactureras y pueden mejorarse con tecnologías inalámbricas como \textit{RFID} para aumentar su agilidad operativa~\cite{stouffer2011ics}.

    \item \textbf{Sistemas de Control Jerárquico:} Finalmente, los sistemas de control jerárquico adoptan una estructura multinivel para manejar operaciones complejas y a gran escala. En este modelo, los controladores locales en el nivel inferior interactúan directamente con los subsistemas, mientras que los niveles superiores se encargan de la supervisión general y la coordinación estratégica. Esta arquitectura permite una gestión eficiente de sistemas industriales complejos, integrando funcionalidades de SCADA y DCS según sea necesario~\cite{stouffer2011ics}. Además, los sistemas jerárquicos facilitan la escalabilidad y la segmentación de tareas, lo que los hace ideales para entornos donde la precisión y la adaptabilidad son prioritarias. Sin embargo, su complejidad puede introducir desafíos adicionales en términos de seguridad y mantenimiento~\cite{khujamatov2021iot}.
\end{itemize}

\newpage
\section{Vulnerabilidades}
Los entornos del Internet Industrial de las Cosas (IIoT) han revolucionado la forma en que las industrias operan, integrando dispositivos inteligentes, sistemas automatizados y plataformas avanzadas. Sin embargo, este crecimiento ha venido acompañado de un aumento exponencial en la cantidad de dispositivos interconectados, lo que implica una mayor exposición a la \textbf{red global}. Este nivel de conectividad ha incrementado significativamente la superficie de ataque, dejando a los sistemas industriales vulnerables frente a amenazas cibernéticas.\\

Una \textbf{vulnerabilidad} puede entenderse como cualquier debilidad inherente en el diseño, la implementación o la configuración de un sistema que puede ser explotada por un atacante para comprometer su integridad, disponibilidad o confidencialidad. En los entornos IIoT, estas vulnerabilidades pueden surgir tanto de los dispositivos individuales, como sensores y actuadores, como de las plataformas que los gestionan, los protocolos de comunicación empleados o las aplicaciones en la nube.\\

Para abordar este tema y poder presentar una exposición clara y organizada de las vulnerabilidades en entornos IIoT utilizaremos como punto de partida la taxonomía propuesta en \textbf{Security Issues in IIoT: A Comprehensive Survey of Attacks on IIoT and Its Countermeasures} \cite{security_issues_iiot} \\

\begin{figure}[H]
    \centering
    \includegraphics[width=0.9\textwidth]{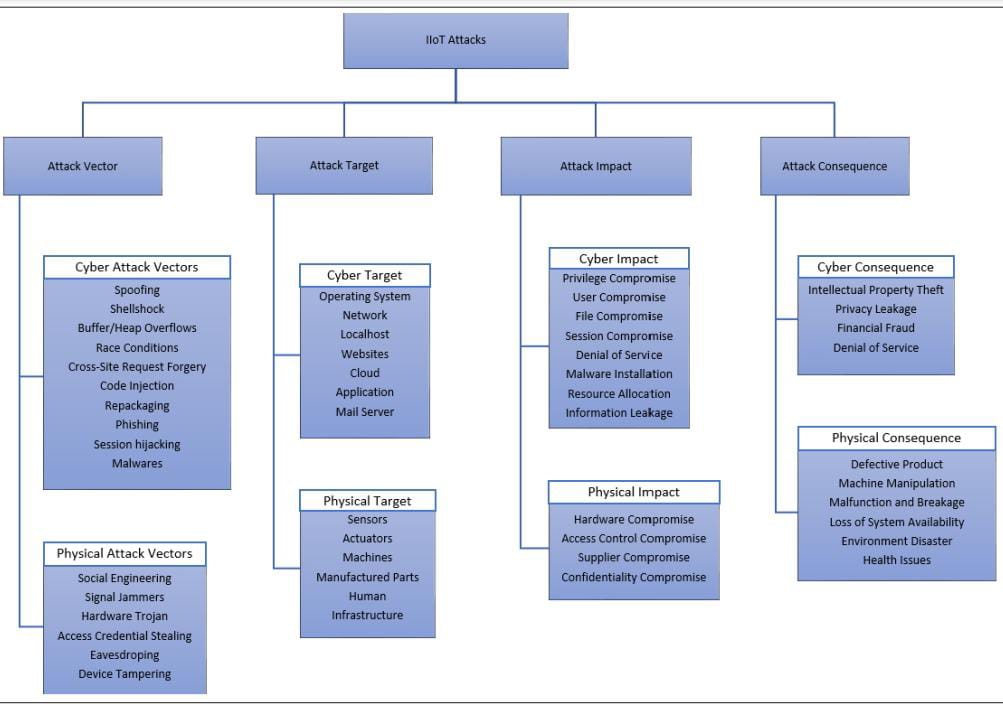}
    \caption{Taxonomia ataques en IIoT. Imagen adaptada de~\cite{security_issues_iiot}.}
    \label{fig:entorno_iot}
\end{figure}
\subsection{Vectores de Ataque}

Un \textbf{vector de ataque} se define como la ruta o método que un adversario utiliza para explotar vulnerabilidades en un sistema, con el objetivo de comprometer su confidencialidad, integridad o disponibilidad. En entornos IIoT, estos vectores son particularmente críticos debido a la integración de sistemas operativos (OT) y tecnológicos (IT), así como a la exposición física de dispositivos en infraestructuras industriales. Según la taxonomía propuesta por Panchal et al.~\cite{security_issues_iiot}, los vectores de ataque en IIoT se clasifican en dos categorías principales: \textit{cibernéticos} y \textit{físicos}. A continuación, se detallan mas en profundiad.\\

\scalebox{1.2}{\textbf{Cibernéticos}}\\

Los vectores de ataque cibernéticos son aquellos que no requieren interacción física con los dispositivos o infraestructuras industriales, centrándose en vulnerabilidades en los sistemas de TI (Tecnología de la Información) y las redes de comunicación. Estos vectores incluyen:

\begin{itemize}
    \item \textbf{Spoofing}: Suplantación de identidad en la que un atacante se hace pasar por un usuario o dispositivo legítimo.
    \item \textbf{Shellshock}: Ejecución remota de código mediante vulnerabilidades en intérpretes de comandos.
    \item \textbf{Buffer/Heap Overflow}: Desbordamiento de memoria que permite la ejecución de código malicioso.
    \item \textbf{Race Condition}: Condiciones de carrera donde múltiples procesos compiten por el acceso a un recurso compartido.
    \item \textbf{CSRF (Cross-Site Request Forgery)}: Envío de solicitudes maliciosas desde un usuario autenticado sin su conocimiento.
    \item \textbf{Inyección de Código (Code Injection)}: Introducción de código malicioso en aplicaciones vulnerables.
    \item \textbf{Reempaquetado (Repackaging)}: Alteración de aplicaciones legítimas para incluir código malicioso.
    \item \textbf{Phishing}: Engaños para obtener información confidencial haciéndose pasar por entidades confiables.
    \item \textbf{Secuestro de Sesión (Session Hijacking)}: Robo de cookies de sesión para tomar el control de una sesión activa.
    \item \textbf{Malware}: Software malicioso, como virus, gusanos, troyanos, rootkits, spyware, botnets y ransomware, diseñado para comprometer sistemas.\\
\end{itemize}

\scalebox{1.2}{\textbf{Físicos}}\\

Los vectores de ataque físicos requieren interacción directa con dispositivos o personas dentro del entorno industrial. Estos vectores incluyen:

\begin{itemize}
    \item \textbf{Jamming de Señales}: Interrupción intencionada de las comunicaciones inalámbricas mediante la emisión de señales interferentes.
    \item \textbf{Troyanos de Hardware }: Modificaciones maliciosas en el hardware durante su diseño o fabricación, activadas bajo condiciones específicas.
    \item \textbf{Robo de Credenciales de Acceso}: Obtención de contraseñas, tarjetas de acceso u otros medios de autenticación legítimos, ya sea mediante espionaje físico (como \textit{shoulder surfing}) o técnicas como el \textit{tailgating}.
    \item \textbf{Escuchas Clandestinas }: Monitoreo no autorizado de comunicaciones privadas para recopilar información confidencial.
    \item \textbf{Manipulación de Dispositivos}: Alteración física de dispositivos, como desactivar mecanismos de seguridad o modificar configuraciones críticas.
    \item \textbf{Ingeniería Social }: Interacción social para engañar a personas y obtener información confidencial o realizar acciones que puedan desencadenar un ataque contra el sistema.\\\\
\end{itemize}

\subsection{Objetivo del Ataque}

En los entornos IIoT, los objetivos de los ataques pueden clasificarse en dos categorías principales: \textbf{Cyber Target (Objetivos Cibernéticos)} y \textbf{Physical Target (Objetivos Físicos)}. Cada uno de estos objetivos abarca componentes críticos dentro de la infraestructura industrial, cuya vulnerabilidad puede comprometer la seguridad, integridad y disponibilidad de los sistemas. A continuación, se detalla cada categoría:\\

\scalebox{1.2}{\textbf{Cibernéticos}}\\

Los objetivos cibernéticos incluyen componentes que forman parte de los sistemas de TI (Tecnología de la Información) y que son esenciales para la operación, gestión y comunicación en los entornos IIoT:

\begin{itemize}
    \item \textbf{Operating System}: Los sistemas operativos en IIoT incluyen sistemas de propósito general y de propósito específico, como RTOS (Real Time Operating System) para dispositivos de baja latencia y sistemas basados en Windows o Linux utilizados en estaciones de trabajo y sistemas SCADA.
    \item \textbf{Network}: Las redes industriales son objetivos frecuentes debido a que muchos protocolos industriales carecen de mecanismos adecuados de autenticación y cifrado.
    \item \textbf{Localhost (Workstation)}: Las estaciones de trabajo pueden ser atacadas para engañar al operador, manipulando alarmas o induciendo acciones perjudiciales.
    \item \textbf{Websites (Web Server)}: Los servidores web se utilizan para alojar información sobre sistemas industriales y compartir datos con las partes interesadas, lo que los hace susceptibles a ataques.
    \item \textbf{Cloud}: La nube, que alberga aplicaciones y datos críticos de las plantas industriales, es un objetivo principal para atacar máquinas virtuales y comprometer la integridad de los datos.
    \item \textbf{Application}: Algunas aplicaciones industriales integradas en dispositivos se utilizan para configurar, probar y recopilar datos, siendo vulnerables a manipulaciones.
    \item \textbf{Mail Server}: Los servidores de correo en la DMZ pueden ser objetivos para obtener credenciales y enviar correos de phishing.\\
\end{itemize}

\scalebox{1.2}{\textbf{Físicos}}\\

Los objetivos físicos incluyen los dispositivos y componentes tangibles que son fundamentales para la operación de los sistemas industriales. Estos objetivos suelen estar expuestos a manipulaciones físicas y otras amenazas específicas:

\begin{itemize}
    \item \textbf{Sensors}: Son esenciales para la monitorización en tiempo real de los procesos industriales.
    \item \textbf{Actuators}: Componentes responsables de convertir señales electrónicas en acciones físicas, como abrir válvulas o accionar motores.
    \item \textbf{Machines}: Equipos diseñados para realizar tareas específicas dentro de procesos industriales, como ensamblaje o transporte de materiales.
    \item \textbf{Manufactured Parts}: Los productos finales o piezas fabricadas pueden ser objetivos de ataques para introducir defectos o vulnerabilidades ocultas.
    \item \textbf{Human}: Los empleados pueden ser víctimas de ataques de ingeniería social, donde se les engaña para revelar información confidencial o realizar acciones perjudiciales.
    \item \textbf{Infrastructure}: Elementos como switches y controladores de energía son críticos para la operación industrial, siendo vulnerables a ataques como el envenenamiento de caché DNS para manipular el flujo de tráfico en la red.\\\\
\end{itemize}

\subsection{Impacto del Ataque}

El impacto de un ataque se define como el efecto o la consecuencia directa que provoca sobre el sistema comprometido. Los ataques en entornos IIoT pueden generar desde modificaciones no deseadas en el comportamiento de los sistemas hasta daños físicos significativos. Estos impactos se dividen en dos categorías principales: \textbf{Impacto Cibernético} e \textbf{Impacto Físico}. A continuación, se detallan los diferentes tipos de impacto dentro de cada categoría:\\

\scalebox{1.2}{\textbf{Cibernéticos}}\\

Los impactos cibernéticos afectan principalmente a los componentes digitales, alterando su funcionalidad, seguridad o disponibilidad. Estos incluyen:

\begin{itemize}
    \item \textbf{Compromiso de Privilegios}: Bypass de las restricciones de privilegios que permite a usuarios o atacantes realizar tareas que pueden afectar el rendimiento o la disponibilidad del sistema.
    \item \textbf{Compromiso de Usuarios}: Acceso no autorizado a la cuenta de un usuario legítimo mediante el robo de credenciales o ataques de fuerza bruta.
    \item \textbf{Compromiso de Archivos}: Modificaciones maliciosas en archivos del sistema, ya sea mediante cambios directos o reempaquetado con certificados robados.
    \item \textbf{Compromiso de Sesiones}: Control de sesiones activas a través de ataques como CSRF o el robo de cookies de sesión.
    \item \textbf{Denegación de Servicio (DoS)}: Saturación de recursos del sistema, denegando el servicio a usuarios legítimos al consumir ancho de banda o capacidad de procesamiento.
    \item \textbf{Instalación de Malware}: Instalación de software malicioso, como troyanos de acceso remoto (RAT), para obtener control remoto sobre el sistema.
    \item \textbf{Bloqueo de Recursos}: Bloqueo de recursos críticos por procesos maliciosos, impidiendo su uso por parte de procesos legítimos.
    \item \textbf{Fuga de Información}: Exposición de datos confidenciales a través de ataques como Man-in-the-Middle (MitM), lo que puede causar pérdidas financieras significativas.\\
\end{itemize}
\newpage
\scalebox{1.2}{\textbf{Físicos}}\\

Los impactos físicos afectan directamente a los componentes materiales del sistema, generando consecuencias que pueden comprometer la infraestructura o la seguridad industrial. Estos incluyen:

\begin{itemize}
    \item \textbf{Compromiso de Hardware}: Malfuncionamiento de componentes de hardware, que puede provocar fallos en procesos críticos o daños materiales significativos.
    \item \textbf{Compromiso de Control de Acceso}: Violación de los mecanismos de control de acceso, permitiendo a atacantes manipular o desactivar dispositivos de manera no autorizada.
    \item \textbf{Compromiso de Proveedores}: Compromiso de la integridad del código fuente o de productos fabricados, lo que puede resultar en pérdida de propiedad intelectual o daños financieros para el proveedor.
    \item \textbf{Compromiso de la Confidencialidad}: Exposición de datos confidenciales, como credenciales o secretos comerciales, a través de canales de comunicación inseguros, comprometiendo la privacidad de la organización.\\\\
\end{itemize}

\subsection{Consecuencias del Ataque}

Las consecuencias de un ataque representan el resultado global que este genera tras su ejecución en un entorno IIoT. Estas consecuencias reflejan lo que el atacante logró tras comprometer el sistema y pueden variar desde la interrupción de servicios hasta daños físicos significativos. Según la taxonomía, las consecuencias de los ataques en IIoT se dividen en dos categorías principales: \textbf{Consecuencias Cibernéticas} y \textbf{Consecuencias Físicas}. A continuación, se detallan las subcategorías de cada una:\\

\scalebox{1.2}{\textbf{Cibernéticas}}\\

Las consecuencias cibernéticas son aquellas que afectan directamente los componentes digitales o virtuales del sistema industrial. Estas incluyen:

\begin{itemize}
    \item \textbf{Robo de Propiedad Intelectual}: Robo de secretos comerciales y datos confidenciales que pueden ser utilizados por competidores para desarrollar estrategias comerciales o productos falsificados.
    \item \textbf{Fuga de Privacidad}: Exposición de información privada de clientes o personas involucradas en los procesos industriales, lo que puede dañar la reputación de la marca y causar grandes pérdidas económicas.
    \item \textbf{Fraude Financiero}: Uso de información financiera y registros de transacciones para cometer fraudes económicos.
    \item \textbf{Denegación de Servicio (DoS)}: Interrupción de servicios legítimos, que puede impedir a las partes interesadas acceder a datos o recursos necesarios.\\
\end{itemize}

\scalebox{1.2}{\textbf{Físicas}}\\

Las consecuencias físicas afectan directamente a los dispositivos, máquinas o infraestructura del entorno industrial, generando daños materiales o riesgos para las personas. Estas incluyen:

\begin{itemize}
    \item \textbf{Producto Defectuoso}: Alteraciones maliciosas en herramientas de fabricación automatizada que conducen a la producción de productos defectuosos.
    \item \textbf{Manipulación de Máquinas}: Modificación del comportamiento de las máquinas sin el conocimiento de los operadores, como el uso de dispositivos IoT para actividades no autorizadas, como minería de criptomonedas.
    \item \textbf{Malfuncionamiento y Rotura}: Malfuncionamiento de dispositivos industriales que puede resultar en daños materiales, como ocurrió con el gusano Stuxnet, que destruyó centrifugadoras en instalaciones nucleares iraníes.
    \item \textbf{Pérdida de Disponibilidad del Sistema}: Ataques que provocan la pérdida de disponibilidad de sistemas o recursos, como el ataque cibernético a la planta de energía ucraniana que desconectó subestaciones durante tres horas.
    \item \textbf{Desastre Ambiental}: Fallos en sistemas de detección de fugas en plantas de petróleo y gas que pueden desencadenar desastres ambientales significativos.
    \item \textbf{Problemas de Salud}: Exposición a productos químicos, gases, radiaciones o malfuncionamientos de maquinaria que pueden poner en peligro la salud de las personas cercanas.
\end{itemize}

\newpage

\subsection{Vulnerabilidades en Protocolos}
En este apartado se muestra a modo de resumen una tabla con las principales vulnerabilidades que presentan los protocolos mas ampliamente utilizados en IIoT mencionados en el apartado 2.3.
\begin{table}[htbp]
\centering
\renewcommand{\arraystretch}{1.0}
\setlength{\tabcolsep}{6pt}
\caption{Resumen de amenazas y vulnerabilidades en protocolos de comunicación}
\label{tab:threats_vulnerabilities}
\begin{tabular}{|c|p{0.7\linewidth}|}
\hline
\rowcolor{gray!25}
\textbf{Protocolo} & \textbf{Amenazas/Vulnerabilidades} \\ \hline

\textbf{CAN} & - Falta de autenticación \cite{bozdal_can_security} \\
             & - Falta de cifrado  \\
             & - Ataques de denegación de servicio (DoS)  \\
             & - Inyección de datos maliciosos  \\
             & - Intercepción de comunicaciones (Eavesdropping) \cite{bozdal_can_security} \\
\hline

\textbf{MQTT} & - Autenticación y cifrado débiles \\
              & - Oscuridad en el puerto  \\
              & - Ataques de denegación de servicio (DoS) \cite{mekala_cybersecurity_iiot} \\
              & - Ataques Man-in-the-Middle (MITM) \cite{nebbione_protocol_security} \\
              & - Ataques de fuerza bruta \\
\hline

\textbf{CoAP} & - Falta de autenticación y autorización \cite{nebbione_protocol_security} \\
              & - Suplantación de identidad (IP spoofing)  \\
              & - Ataques Man-in-the-Middle (MITM)  \\
              & - Ataques de denegación de servicio (DoS)  \\
\hline

\textbf{DDS} & - Cifrado y autorización deficientes  \\
             & - Ataques de denegación de servicio (DoS) \\
             & - Ataques Man-in-the-Middle (MITM) \cite{nebbione_protocol_security} \\
\hline

\textbf{AMQP} & - Autenticación y cifrado deficientes  \\
              & - Ataques de denegación de servicio (DoS) \cite{nebbione_protocol_security} \\
              & - Inyección de datos maliciosos \\
              & - Ataques Man-in-the-Middle (MITM)  \\
              & - Secuestro de tráfico \cite{mekala_cybersecurity_iiot} \\
              & - Ejecución remota de código \ \\
\hline

\textbf{Fieldbus} & - Cifrado y autenticación débiles \cite{thomesse_fieldbus} \\
                  & - Ataques de denegación de servicio (DoS)  \\
                  & - Desbordamiento de búfer \cite{mekala_cybersecurity_iiot} \\
                  & - Inyección de comandos maliciosos  \\
                  & - Ataques Man-in-the-Middle (MITM)  \\
                  & - Intercepción de comunicaciones (Eavesdropping)  \\
\hline

\textbf{MODBUS/TCP} & - Falta de autenticación, cifrado y control de acceso \cite{rahman_modbus_security} \\
                    & - Falta de integridad de datos \\
                    & - Alteración de datos  \\
                    & - Inyección de datos maliciosos \\
                    & - Ataques de denegación de servicio (DoS)  \\
                    & - Ataques de repetición (Replay attack)  \\
                    & - Ataques Man-in-the-Middle (MITM)  \\
\hline

\textbf{WirelessHART} & - Autenticación y cifrado débiles \cite{devan_wirelesshart} \\
                      & - Inyección de paquetes maliciosos  \\
                      & - Ataques de denegación de servicio (DoS)  \\
                      & - Ataques Man-in-the-Middle (MITM)  \\
                      & - Ataques de interferencia (Jamming)  \\
                      & - Suplantación de identidad y espionaje (Spoofing y Eavesdropping)  \\
\hline

\textbf{NB-IoT} & - Inyección de paquetes maliciosos \cite{mekala_cybersecurity_iiot} \\
                & - Desbordamiento de búfer  \\
                & - Ataques de denegación de servicio (DoS) \cite{mekala_cybersecurity_iiot} \\
                & - Ataques de interferencia (Jamming)  \\
                & - Ataques de repetición (Replay)  \\
\hline
\end{tabular}
\end{table}

\newpage

\section{Ataques}
\subsection{Fases de Ataque}
A la hora de enfrentar los problemas de seguridad en un sistema es crucial comprender los patrones generales que suelen
seguir los ataques a estos mismos, si lo que queremos es desarrollar herramientas verdaderamente capaces de hacerles
frente. En el contexto de los ataques a infraestructuras de Internet de las Cosas Industrial (IIoT), los patrones de ataque
suelen seguir una secuencia estructurada que puede variar ligeramente dependiendo del marco de referencia utilizado.
Algunos de los marcos más conocidos son el Ciclo de Vida del Ataque de Lockheed Martin (CKC) \cite{lockheed_martin_ckc}, el Ciclo de Vida del
Ataque de Mandiant (MALC) \cite{mandiant_malc} y el marco ATT\&CK de la MITRE Corporation \cite{mitre_att&ck}. Aunque estos marcos fueron originalmente
diseñados para entornos de Tecnologías de la Información (IT), se han adaptado para abordar ataques en entornos de
Sistemas de Control Industrial (ICS) y, en menor medida, en IIoT. Sin embargo, debido a la naturaleza híbrida de IIoT,
que combina tanto tecnologías de Operaciones Tecnológicas (OT) como nuevas tecnologías de IT, es necesario adaptar estos
marcos para capturar mejor la complejidad de los ataques en este tipo de infraestructuras \cite{xiiotid}.

\begin{itemize}
    \item \textbf{Reconocimiento:} La fase de reconocimiento es el punto de partida de cualquier ataque. En esta fase, el atacante busca información sobre el objetivo, identifica vulnerabilidades y selecciona el método de ataque más adecuado. Las técnicas comunes incluyen el escaneo genérico para identificar puertos abiertos, sistema operativo y servicios disponibles en el objetivo utilizando herramientas como Nmap; el escaneo de vulnerabilidades para buscar vulnerabilidades conocidas y malas configuraciones en el sistema objetivo utilizando bases de datos como CVE; el fuzzing para enviar datos aleatorios o semiválidos a un sistema y detectar errores y excepciones; y el descubrimiento de recursos para identificar recursos disponibles en un sistema IIoT, como sensores y actuadores, mediante solicitudes específicas. Esta fase es crucial para el atacante, ya que proporciona la información necesaria para planificar el ataque. Detectar a tiempo actividades de reconocimiento puede ayudar a prevenir ataques incipientes.
    
    \item \textbf{Arma:} La fase de arma, también conocida como compromiso inicial, es donde el atacante obtiene acceso al sistema objetivo. Las técnicas comunes incluyen el ataque de fuerza bruta, que consiste en intentos repetidos de acceso utilizando combinaciones de nombre de usuario y contraseña; el ataque de diccionario, que utiliza listas de palabras para intentar adivinar contraseñas; y el insider malicioso, que es un empleado con acceso legítimo que actúa con intenciones maliciosas. Esta fase es crítica para el éxito del ataque. 
    
    \item \textbf{Explotación:} En esta fase, el atacante explota una vulnerabilidad para obtener acceso más profundo al sistema. Las técnicas comunes incluyen el shell inverso, que crea una conexión inversa desde el sistema comprometido al sistema del atacante; y el ataque de hombre en el medio (MitM), que intercepta la comunicación entre dos puntos para inyectar datos maliciosos. La explotación puede ser rápida y difícil de detectar.
    
    \item \textbf{Movimiento Lateral:} El objetivo de esta fase es expandir el acceso dentro de la red y comprometer más sistemas. Las técnicas comunes incluyen la suscripción a broker MQTT para obtener información de dispositivos físicos; la lectura de registros Modbus para leer y descubrir registros en dispositivos PLC; y el ataque de retransmisión TCP, que utiliza técnicas de pivoteo para moverse entre segmentos de red. El movimiento lateral puede ser lento y discreto. La segmentación de la red y la monitorización de accesos internos pueden limitar la propagación del ataque.
    
    \item \textbf{Comando y Control (C\&C):} Esta fase implica la creación de un canal de comunicación entre el sistema comprometido y el servidor del atacante para recibir comandos y enviar datos. Las técnicas comunes incluyen el uso de túnel DNS para establecer una comunicación discreta. El C\&C es esencial para el control del ataque. La implementación de firewalls pueden interrumpir estos canales.
    
    \item \textbf{Exfiltración:} La exfiltración es el proceso de extraer información sensible del sistema comprometido. Las técnicas comunes incluyen la compresión y obfuscación de datos para evitar la detección durante la transferencia de datos. La exfiltración puede ser difícil de detectar. Encriptar los datos y monitorizar el tráfico saliente pueden ayudar a proteger la información.
    
    \item \textbf{Alteración:} Esta fase implica la manipulación de datos para afectar la integridad de la información. Las técnicas comunes incluyen la inyección de datos falsos en la nube para afectar análisis de datos y el envío de notificaciones falsas a operadores. La alteración puede tener consecuencias graves en la toma de decisiones.
    
    \item \textbf{Cripto-Ransomware:} En esta fase, el atacante inyecta malware para cifrar datos y exigir un rescate en criptomonedas. Las técnicas comunes incluyen el cifrado de archivos críticos hasta que se pague el rescate. El cripto-ransomware puede ser devastador. Crear copias de seguridad regulares pueden mitigar el impacto.
    
    \item \textbf{Denegación de Servicio Extorsivo (RDoS):} Esta fase implica amenazar con un ataque de denegación de servicio a menos que se pague un rescate. 
\end{itemize}
\newpage

\subsection{Recopilacion Ataques}
Para ilustrar todo lo expuesto anteriormente se va a elaborar una tabla con ataques reales que han afectado a infraestructura IIoT y clasificados en base a la taxonomia expuesta en el apartado anteriror.

\begin{center}
    \begin{longtable}{|p{3.5cm}|p{3.5cm}|p{3.5cm}|p{3.5cm}|p{3.5cm}|}
        \hline
        \rowcolor{gray!25} 
        \textbf{Nombre} & 
        \textbf{Vector} & 
        \textbf{Objetivo} & 
        \textbf{Impacto} & 
        \textbf{Consecuencia} \\ 
        \hline
        \endfirsthead

        \multicolumn{5}{c}{{\tablename\ \thetable{} -- continuación de la página anterior}} \\ 
        \hline
        \rowcolor{gray!25} 
        \textbf{Nombre} & 
        \textbf{Vector} & 
        \textbf{Objetivo} & 
        \textbf{Impacto} & 
        \textbf{Consecuencia} \\ 
        \hline
        \endhead

        \hline
        \multicolumn{5}{r}{{Continuará en la siguiente página}} \\ 
        \hline
        \endfoot

        \hline
        \endlastfoot

         \textbf{1. Stuxnet} \cite{denning2012stuxnet} & Cibernético (Malware) & \textit{Machines / Actuators} (Controladores de centrifugadoras) & \textit{Compromiso de Hardware} (Modificación de operaciones críticas) & \textit{Malfuncionamiento y Rotura} (Daños físicos en centrifugadoras) \\ \hline

    \textbf{2. Troyano de Hardware} \cite{alkhalifa2015trojan} & Físico (Inserción de componentes maliciosos) & \textit{Infrastructure / Machines} (Placa base de un dispositivo industrial) & \textit{Compromiso de Hardware} (Dispositivo inseguro) & \textit{Producto Defectuoso} (Fallos o puertas traseras) \\ \hline

    \textbf{3. Ataque DDoS a un SCADA} \cite{zhukabayeva2024ddos} & Cibernético (DoS / DDoS) & \textit{Network / SCADA} & \textit{Denegación de Servicio} (Saturación de recursos) & \textit{Pérdida de Disponibilidad} (Interrupción industrial) \\ \hline

    \textbf{4. Manipulación de Sensor} \cite{zhang2024sensor} & Físico (Alteración directa) & \textit{Sensors} & \textit{Compromiso de Control de Acceso} (Acceso no autorizado) & \textit{Producto Defectuoso} (Errores en la producción) \\ \hline

    \textbf{5. Ransomware en la Nube} \cite{alkhalifa2022ransomware} & Cibernético (Malware) & \textit{Cloud / Application} & \textit{Instalación de Malware} (Cifrado de archivos) & \textit{Denegación de Servicio} (Bloqueo de acceso a datos) \\ \hline

    \textbf{6. Desbordamiento de Búfer en un PLC} \cite{alkhalifa2023buffer} & Cibernético (Buffer Overflow) & \textit{Actuators / Machines} & \textit{Compromiso de Hardware} (Ejecución de código malicioso) & \textit{Malfuncionamiento y Rotura} (Daños físicos) \\ \hline

        \textbf{7. Jamming de Señales} & Físico (Jamming) & \textit{Infrastructure (Antenas, APs)} & \textit{Bloqueo de Recursos} (Red inalámbrica inutilizada) & \textit{Pérdida de Disponibilidad} (Interrupción de comunicaciones) \\ \hline

        \textbf{8. Dispositivo USB Infectado} & Físico (Interacción directa) & \textit{Localhost (Workstation)} & \textit{Instalación de Malware} (Troyano o keylogger) & \textit{Fuga de Privacidad} (Exfiltración de datos) \\ \hline

        \textbf{9. Tailgating (Ingeniería Social)} & Físico (Acceso físico no autorizado) & \textit{Human / Infrastructure} & \textit{Compromiso de Control de Acceso} (Ingreso sin permisos) & \textit{Robo de Información} (Acceso a red interna) \\ \hline

       \textbf{10. Reempaquetado de Aplicaciones IIoT} \cite{alkhalifa2023buffer} & Cibernético (Repackaging) & \textit{Application / OS} & \textit{Instalación de Malware} (Código espía embebido) & \textit{Fuga de Privacidad} (Exfiltración de datos) \\ \hline

        \textbf{11. Secuestro de Sesión Web SCADA} \cite{keepcoding2021hijacking} & Cibernético (Session Hijacking) & \textit{Web Server} & \textit{Compromiso de Sesiones} (Control de sesión de operador) & \textit{Denegación de Servicio} (Cierre de procesos) \\ \hline

        \textbf{12. CSRF en Gestión IIoT} \cite{alkhalifa2022ransomware} & Cibernético (CSRF) & \textit{Websites / Application} & \textit{Compromiso de Sesiones} (Acciones no autorizadas) & \textit{Fraude Financiero} (Manipulación de parámetros) \\ \hline

        \textbf{13. Alteración de Firmware en IIoT} \cite{chaudhary2023ddos} & Físico (Troyanos de Hardware) & \textit{Sensors / Actuators} & \textit{Compromiso de Hardware} (Modificación de firmware) & \textit{Producto Defectuoso} (Resultados alterados) \\ \hline

        \textbf{14. Race Condition en OS Embebido} \cite{alkhalifa2023buffer} & Cibernético (Race Condition) & \textit{Operating System} & \textit{Compromiso de Archivos} (Corrupción de archivos) & \textit{Denegación de Servicio} (Fallo masivo del sistema) \\ \hline

        \textbf{15. Robo de Credenciales con \textit{Shoulder Surfing}} \cite{lasexta2022shouldersurfing} & Físico (Robo de Credenciales de Acceso) & \textit{Physical Target: Human} & \textit{Compromiso de Usuarios} (Acceso posterior a la red con credenciales legítimas) & \textit{Fraude Financiero / Fuga de Privacidad} (Uso indebido de cuentas con privilegios) \\ \hline

        \textbf{16. Falsificación de Certificados de Software IIoT} \cite{keyfactor2024trust} & Cibernético (Spoofing) & \textit{Cyber Target: Application / Operating System} & \textit{Compromiso de Archivos} (Ejecutables maliciosos parecen legítimos) & \textit{Robo de Propiedad Intelectual o Fuga de Privacidad} (Distribución de software comprometido que filtra datos) \\ \hline

        \textbf{17. Ataque de Cadena de Suministro (SolarWinds-style en IIoT)} \cite{sanchez2023cybersecurity} & Cibernético (Reempaquetado/Malware) & \textit{Application/Cloud} (Actualizaciones de software legítimas comprometidas) & \textit{Compromiso de Archivos} (Inserción de código malicioso en actualizaciones) & \textit{Robo de Propiedad Intelectual} (Acceso a diseños industriales o datos sensibles) \\ \hline

    \textbf{18. Envenenamiento de Modelos de ML/AI} \cite{dunn2020robustness} & Cibernético (Inyección de Código) & \textit{Application} (Sistemas de mantenimiento predictivo basados en IA) & \textit{Compromiso de Archivos} (Alteración de conjuntos de datos de entrenamiento) & \textit{Producto Defectuoso} (Decisiones erróneas en líneas de producción) \\ \hline

    \textbf{19. Ataque de Canal Lateral (Side-Channel)} \cite{ferrag2022edgeiiotset} & Físico (Escuchas Clandestinas) & \textit{Sensors/Actuators} (Dispositivos con emisiones electromagnéticas no protegidas) & \textit{Fuga de Información} (Extracción de claves criptográficas mediante análisis de energía) & \textit{Robo de Propiedad Intelectual} (Acceso a algoritmos propietarios de control) \\ \hline

    \textbf{20. Explotación de Firmware no Parcheado (Ej. VxWorks)} \cite{ullah2023magru} & Cibernético (Buffer Overflow) & \textit{Operating System} (RTOS en dispositivos edge como PLCs) & \textit{Compromiso de Privilegios} (Ejecución remota de código) & \textit{Manipulación de Máquinas} (Parálisis de líneas de ensamblaje) \\ \hline

    \textbf{21. DNS Spoofing en Redes OT} \cite{elias2022dns} & Cibernético (Spoofing) & \textit{Network} (Servidores DNS internos en redes industriales) & \textit{Fuga de Información} (Redirección de tráfico a servidores maliciosos) & \textit{Robo de Propiedad Intelectual} (Interceptación de datos de procesos industriales) \\ \hline

    \textbf{22. Cryptojacking en Dispositivos Edge} \cite{saadouni2023cryptojacking} & Cibernético (Malware) & \textit{Machines} (Gateways IIoT con capacidad de procesamiento) & \textit{Bloqueo de Recursos} (Consumo de CPU para minería de criptomonedas) & \textit{Pérdida de Disponibilidad} (Retrasos críticos en tiempo real en líneas de producción) \\ \hline

    \textbf{23. Ataque de Falsa Inyección de Datos (False Data Injection)} \cite{himeur2023federated} & Cibernético (Spoofing) & \textit{Sensors} (Sensores de temperatura/presión en oleoductos) & \textit{Compromiso de Archivos} (Manipulación de lecturas enviadas al SCADA) & \textit{Desastre Ambiental} (Sobrepresión no detectada en tuberías, causando fugas) \\ \hline

    \textbf{24. Ataque de Agotamiento de Batería (Battery Drain)} \cite{javeed2023battery} & Físico (Jamming de Señales) & \textit{Infrastructure} (Dispositivos IIoT inalámbricos con batería limitada) & \textit{Bloqueo de Recursos} (Interrupción de comunicaciones por agotamiento energético) & \textit{Pérdida de Disponibilidad} (Caída de redes de sensores en plantas remotas) \\ \hline

    \textbf{25. Exploit de Protocolos Legacy (Ej. Modbus)} \cite{chaudhary2023ddos} & Cibernético (Inyección de Código) & \textit{Network} (Protocolos sin autenticación como Modbus TCP) & \textit{Compromiso de Control de Acceso} (Comandos no autorizados a PLCs) & \textit{Malfuncionamiento y Rotura} (Parada abrupta de motores industriales) \\ \hline

    \textbf{26. Ataque a Sistemas de Edge Computing} \cite{ferrag2022edgecomputing} & Cibernético (Race Condition) & \textit{Cloud/Edge Nodes} (Nodos de procesamiento local en fábricas) & \textit{Compromiso de Archivos} (Corrupción de datos en tiempo real) & \textit{Producto Defectuoso} (Errores en inspecciones visuales automatizadas) \\ \hline

    \end{longtable}
\end{center}

\noindent Cada uno de estos ataques ilustra cómo se pueden combinar diferentes tipos de vectores (cibernéticos o físicos) con objetivos tanto cibernéticos (p.ej. sistemas operativos, redes, servidores de aplicaciones) como físicos (sensores, actuadores, máquinas), generando impactos que abarcan desde la simple alteración de archivos o sesiones hasta la destrucción de hardware industrial o la producción de bienes defectuosos, con consecuencias igualmente diversas (desde el robo de información confidencial hasta daños materiales y riesgos para la salud y el medioambiente).
\section{Contramedidas de Seguridad}
Los desafíos de ciberseguridad en el entorno industrial (IIoT) son significativamente diferentes y más complejos en comparación con los del IoT orientado al consumidor. Mientras que en el IoT convencional los dispositivos suelen conectarse directamente a Internet para proporcionar o ejecutar sus funciones, en el IIoT existe una fuerte interconexión entre dispositivos de campo, controladores y servidores centralizados, donde una gran parte del procesamiento de datos ocurre dentro de redes locales.\\

La conexión a Internet y los servicios en la nube en el entorno IIoT se utiliza principalmente para mejorar las capacidades de procesamiento local mediante la monitorización avanzada y la optimización de procesos. En este contexto, los requisitos de seguridad más críticos en el IIoT son la disponibilidad y la integridad, aspectos que marcan una clara diferencia con otros dominios.\\

Además, el modelo de atacante en el ámbito del IIoT es un componente vital para clasificar y determinar las amenazas y riesgos asociados. Dado el entorno único del IIoT, caracterizado por procesos críticos, componentes operativos de larga duración, altas demandas de conectividad, un gran número de dispositivos, confidencialidad de datos, errores humanos y posibles sabotajes , se requieren soluciones robustas y específicas para mitigar las amenazas cibernéticas y sus vulnerabilidades asociadas.\\

A continuación, se presentan las principales contramedidas de seguridad propuestas para abordar estos desafíos en el ámbito del IIoT.\\\\

\subsection{IDS para IIoT}
Los Sistemas de Detección de Intrusos (IDS) son esenciales para detectar tráfico malicioso en entornos IIoT, actuando como una capa de defensa secundaria más allá de los firewalls tradicionales. Las soluciones IDS suelen dividirse en dos categorías: basadas en firmas (coincidencia de patrones contra amenazas conocidas) y basadas en anomalías (detección de desviaciones del comportamiento normal). Los enfoques híbridos que combinan ambos métodos son cada vez más populares debido a su capacidad para identificar tanto ataques conocidos como desconocidos.\\

Por ejemplo, en \cite{kirupakar2019intrusion} se propuso un IDS híbrido liviano que utiliza nodos basados en agentes, aprovechando metadatos del sistema y parámetros de contexto para monitorear las pasarelas IIoT. Sin embargo, este enfoque puede volverse computacionalmente costoso. De manera similar, en \cite{yao2019hybrid} se introdujo un IDS híbrido basado en aprendizaje automático (ML) para IIoT en el borde de la red, utilizando gradient boosting para los nodos de borde y aprendizaje profundo para los nodos maestros. Aunque es eficaz para dispositivos con recursos limitados, presenta limitaciones con conjuntos de datos pequeños. Finalmente, en \cite{nyasore2020modbus} se presentó un IDS basado en inspección profunda de paquetes para detectar vulnerabilidades en Modbus/TCP; no obstante, su enfoque en un solo protocolo limita su aplicabilidad general.\\\\

\subsection{Aprendizaje Automático}
Las técnicas de Aprendizaje Automático (ML) son fundamentales para la detección de ataques en entornos IIoT debido a la naturaleza dinámica de estos sistemas y las limitaciones de los IDS tradicionales. Métodos convencionales, como los IDS basados en firmas o reglas, tienen dificultades para adaptarse a amenazas emergentes y cambios en el sistema. En cambio, los IDS basados en ML pueden identificar patrones complejos de anomalías y adaptarse a intrusiones desconocidas.\\

Por ejemplo, en \cite{abdelbasset2020deepifs} se introdujo un modelo de aprendizaje profundo llamado Deep-IFS, que combinó unidades recurrentes cerradas y capas de atención múltiple para detectar intrusiones en entornos de computación en la niebla, demostrando un rendimiento superior en los conjuntos de datos BotIoT y UNSW-NB15. Estos enfoques destacan el potencial del ML para mejorar la precisión de detección y la escalabilidad en entornos IIoT con recursos limitados.\\

Sin embargo, persisten desafíos, como la dependencia de grandes conjuntos de datos etiquetados y el costo computacional. Por ejemplo, en \cite{koroniotis2020forensic} se propuso un marco basado en optimización de enjambre de partículas para ajustar hiperparámetros de redes neuronales profundas, alcanzando una precisión del 99.90\% en Bot-IoT pero requiriendo recursos computacionales significativos. Otro estudio en \cite{koroniotis2020forensic} desarrolló un sistema de detección de anomalías utilizando autoencoders y análisis de componentes principales, el cual enfrentó limitaciones en el manejo de relaciones no lineales entre características. Estos ejemplos subrayan la necesidad de modelos de ML ligeros y adaptativos diseñados para las restricciones del IIoT.\\

Los enfoques híbridos, como la combinación de ML con inspección profunda de paquetes (DPI), muestran potencial para abordar vulnerabilidades específicas de protocolos. Por ejemplo, en \cite{nyasore2020modbus} se propuso un IDS con DPI habilitado para Modbus/TCP, aunque su aplicabilidad se limitó a un solo protocolo. Trabajos futuros deberían centrarse en desarrollar marcos de ML robustos y multimodales que integren datos de sensores y protocolos diversos para garantizar una detección integral de amenazas en entornos IIoT.\\

\begin{figure}[H]
    \centering
    \includegraphics[width=0.9\textwidth]{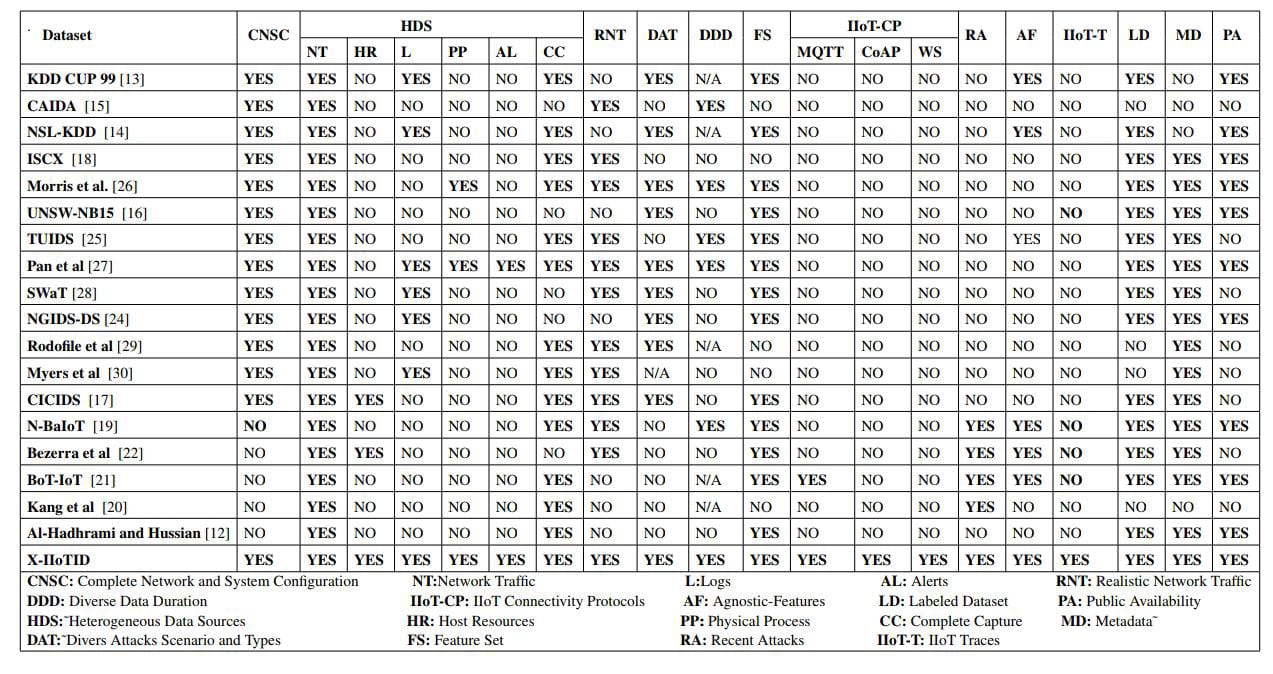}
    \caption{Tabla de comparación de datasets de intrusiones en IIoT~\cite{security_issues_iiot}.}
    \label{fig:entorno_iot}
\end{figure}

\subsection{Protección de redes SCADA}

Las redes SCADA (\textit{Supervisory Control and Data Acquisition}) son esenciales para controlar y monitorear procesos industriales, pero su arquitectura tradicional las hace vulnerables a ciberataques. Una red SCADA típica incluye una red de control (con RTUs, PLCs y sensores), una infraestructura de comunicación (como \textit{fieldbus} o Modbus) y una red de procesos (con servidores y HMI). Cada componente expone riesgos: la infraestructura de comunicación puede sufrir ataques DoS o \textit{jamming}, mientras que las redes inalámbricas de sensores enfrentan amenazas como inyección de datos falsos o replicación de nodos.\\

Para proteger estas redes, se han desarrollado soluciones como \textit{middlewares} resistentes, sistemas de detección de intrusiones (IDS) y técnicas de \textit{machine learning} (ML). Un ejemplo es el IDS basado en ML propuesto en \cite{zolanvari2019ml}, que utiliza algoritmos como \textit{random forest} para detectar comportamientos anormales y alertar a los operadores a través de una interfaz gráfica. Otra aproximación es el modelo de detección de ataques basado en aprendizaje \textit{ensemble} (como \textit{random subspace-random tree}) presentado en \cite{hassan2020trustworthiness}, que combina múltiples clasificadores para mejorar la precisión y reducir el sobreajuste. Estos sistemas permiten identificar amenazas sin comprometer el rendimiento de la red.\\

Además, se han propuesto \textit{middlewares} basados en multiagentes para monitorear y coordinar la comunicación entre componentes, como el \textit{framework} resiliente en \cite{januario2016scada}. Estos sistemas adaptativos ayudan a mitigar amenazas mediante mecanismos de resiliencia y conciencia contextual. En resumen, la seguridad de las redes SCADA requiere soluciones integradas que combinen ML, IDS y arquitecturas flexibles para enfrentar amenazas en tiempo real.\\\\

\subsection{Otras}

Las tecnologías emergentes, como el cifrado adaptativo y los contratos inteligentes, son cruciales para proteger datos en entornos heterogéneos, como se discute en \cite{serror2020iiotsecurity}, donde se proponen soluciones como re-cifrado y cifrado parcial para garantizar confidencialidad de extremo a extremo.El blockchain emerge como una solución para asegurar la integridad y trazabilidad en IIoT. En \cite{kumar2019secblockedge}, se presenta un marco blockchain-edge que aborda amenazas en capas locales (sensores), de borde (procesamiento) y globales (almacenamiento en la nube).\\

En la capa local, se mitigan ataques como inyección de datos mediante firmas criptográficas. En la capa de borde, se abordan amenazas de virtualización mediante aislamiento de políticas de seguridad. En la capa global, se enfrentan ataques API y DoS/DDoS con cifrado homomorfo y encriptación basada en atributos. Finalmente, en la capa de ledger, se protege contra ataques Sybil y de claves privadas mediante cadenas híbridas y cierre de código. Estos enfoques integran blockchain y cifrado para fortalecer la seguridad en todo el ecosistema IIoT.\\

\newpage
\printbibliography

\end{document}